\documentclass[aps,pre,reprint,showpacs,superscriptaddress,groupaddress,showkeys,floatfix,byrevtex3]{revtex4-1}
\usepackage{graphicx}
\usepackage{dcolumn}
\usepackage{bm}
\usepackage{srctex}
\usepackage{stackrel}
\usepackage{amsmath}
\usepackage{amsfonts}
\usepackage{amssymb}
\usepackage[dvipsnames]{xcolor}
\usepackage{cancel}
\usepackage{soul}

\begin{document}
%%%%%%%%%%%%%%%%%%%%%%%%%%%%%%%%%%%

\title{Anomalous diffusion of scaled Brownian tracers}

\author{Francisco J. Sevilla}
\email[]{fjsevilla@fisica.unam.mx}
\thanks{corresponding author}
\affiliation{Instituto de F\'isica, Universidad Nacional Aut\'onoma de M\'exico,
Apdo.\ Postal 20-364, 01000, Ciudad de M\'exico, M\'exico}

\author{Adriano \surname{Vald\'es-G\'omez}}
\affiliation{AI Factory at BBVA M\'exico}
\affiliation{Facultad de Ciencias, Universidad Nacional Autónoma de México, Alcaldía Coyoacán,
C.P. 04510 Ciudad Universitaria, Ciudad de México, México}

\author{Alexis Torres-Carbajal}
\affiliation{Instituto de F\'isica, Universidad Nacional Aut\'onoma de M\'exico,
Apdo.\ Postal 20-364, 01000, Ciudad de M\'exico, M\'exico}

\date{\today}

\begin{abstract}
A model for anomalous transport of tracer particles diffusing in complex media in two dimensions is proposed. The model takes into account the characteristics of persistent motion that active bath transfer to the tracer, thus the model proposed in here extends \emph{active Brownian motion}, for which the stochastic dynamics of the orientation of the propelling force is described by \emph{scale Brownian motion} (sBm), identified by a the time dependent diffusivity of the form $D_\beta\propto t^{\beta-1}$, $\beta>0$. If $\beta\neq1$, sBm is highly non-stationary and suitable to describe such a non-equilibrium dynamics induced by complex media. In this paper we provide analytical calculations and computer simulations to show that genuine \emph{anomalous diffusion} emerge in the long-time regime, with a time scaling of the mean square displacement $t^{2-\beta}$, while ballistic transport $t^2$, characteristic of persistent motion, is found in the short-time one. An analysis of the time dependence of the kurtosis, and  \emph{intermediate scattering function} of the positions distribution, as well as the propulsion auto-correlation function which define the \emph{effective persistence time} are provided. 
\end{abstract}

\maketitle

\section{Introduction}
\emph{Anomalous diffusion} refers to those transport processes whose mean-squared displacement (MSD) scales asymptotically as a power law with time, i.e.,
\begin{equation}
\langle\boldsymbol{x}^2(t)\bigr\rangle\sim t^{\beta}
\end{equation}
with $\beta>0$. The term \emph{subdiffusion} is used for the cases $0<\beta<1$, while \emph{superdiffusion} denotes those for which $\beta>1$. \emph{Normal diffusion} is the term left for the traditional diffusion of a Brownian particle for which $\beta=1$.   

Depending on the specific mechanism that leads to anomalous transport of tracer particles in complex environments, if known, a specific mathematical model can be devised to describe the power-law time dependence of the MSD. Commonly, if the origin of the anomalous behavior is unclear, predictions of the time-scaling of the MSD given by different models are fitted to the the experimental results to obtain the values of the relevant parameters of the model. A standard approach to model the effects of complex environments on the tracer motion is to implicitly incorporate them, in the form of complex noise, into a stochastic differential equations model (Langevin-like models) \cite{WuPhysRevLett2000,KurtulduPNAS2011,KanazawaNature2020}. 

Recently, anomalous diffusion of tracer particles in complex media, particularly in nonequilibrium environments made of active particles (active baths) \cite{WuPhysRevLett2000,KurtulduPNAS2011,KanazawaNature2020,ThapaJChemPhys2019,CaoMolPhys2020} or of viscoleastic fluid \cite{VandebroekPRE2015,VandebroekJStatPhys2017,YuanPhysChemChemPhys2019}, has been observed. In these cases, the transport properties of the tracer particle mimics the ones of active motion, mainly, they exhibit highly \emph{persistent} motion. Active or propelled particles is the name for those agents that locally transform, through complex mechanisms, the energy adsorbed from the environment into a variety forms of locomotion. Active Brownian motion is a simple stochastic model for active motion and is based in the overdamped motion of a Brownian particle subject to propelling force $v_0\hat{\boldsymbol{v}}(t)$, i.e.,
\begin{equation}
\frac{d}{dt}\boldsymbol{x}(t)=v_0\hat{\boldsymbol{v}}(t),
\end{equation}
where $v_0$ is the constant propulsion speed, and $\hat{\boldsymbol{v}}(t)=\bigl(\cos\varphi(t),\sin\varphi(t)\bigr)$ the direction of propulsion, randomized by the fluctuations of the propelling mechanism, in the simplest case,these are modeled by rotational Brownian motion. 

In this paper we consider the case for which the stochastic dynamics of the propelling direction of active Brownian particles is modeled by scaled Brownian motion. In contrast to the results reported in \cite{Gomez-SolanoJStatMech2020}, where the dynamics of the propelling direction was modeled by fractional Brownian motion, in here we show the emergence of genuine anomalous diffusion. 

\emph{Scaled} or \emph{geometric Brownian motio}n, is a nonstationary and nonergodic stochastic process $\chi_{\beta}(t)$ whose time evolution is given by the Langevin-like stochastic differential equation \cite{LimPhysRevE2002,ThielPhysRevE2014b,JeonPhysChemChemPhys2014,SafdariJPhysA2015}
\begin{equation}\label{sBm-Langevin}
\frac{d}{dt}\chi_{\beta}(t)=\sqrt{2\beta K_\beta t^{\beta-1}}\xi(t),    
\end{equation}
where $\xi(t)$is Gaussian white noise, i.e., $\langle\xi(t)\rangle=0$ and $\langle\xi(t)\xi(s)\rangle=\delta(t-s)$.
Scaled Brownian motion has been considered in variety of physical and nonphysical processes and it naturally provides a seemingly adequate description in the case of unbounded diffusion of anomalous diffusion since for $0<\beta<1$, the amplitude of fluctuations in Eq. \eqref{sBm-Langevin} extinguishes with time giving rise to subdiffusion $\bigl\langle\chi^2(t)\bigr\rangle\propto t^{\beta}$, while superdiffusion is observed if $1<\beta$ when the fluctuations grows without limit with time.
\begin{figure*}
\includegraphics[width=0.2\textwidth]{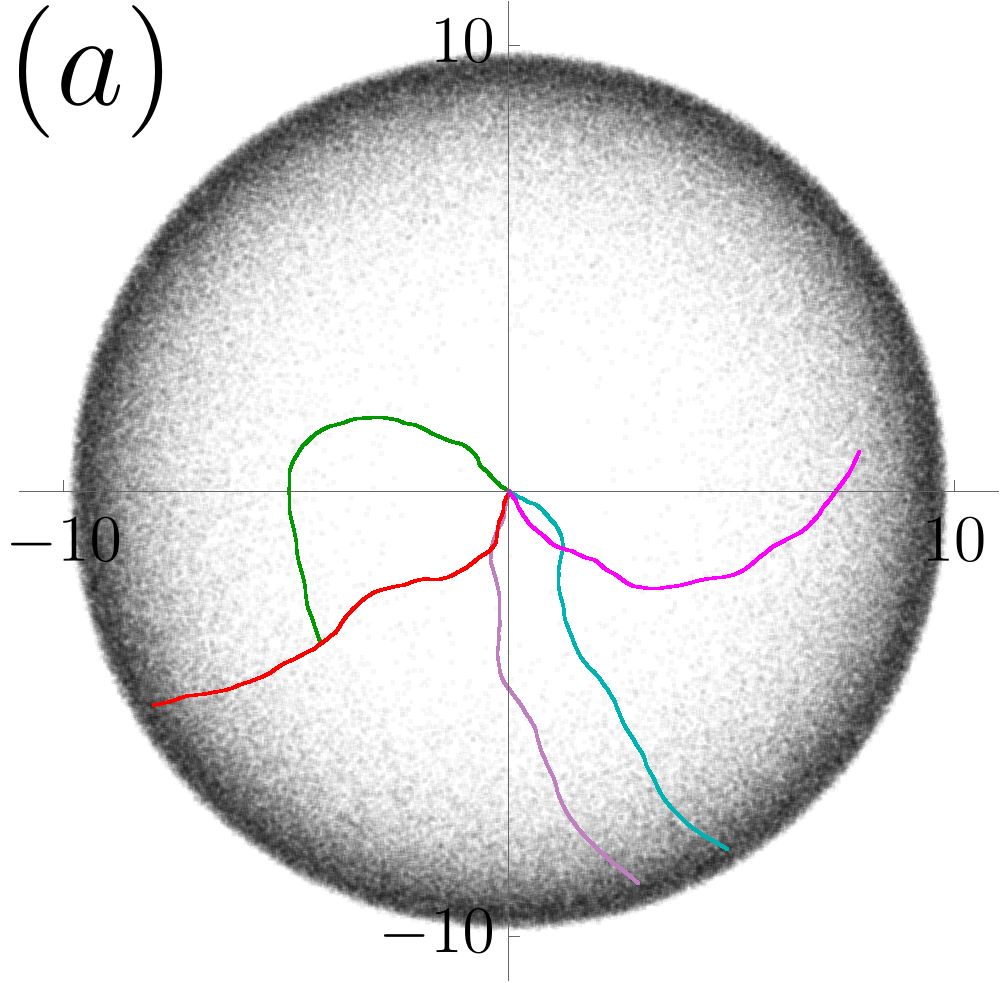}\includegraphics[width=0.2\textwidth]{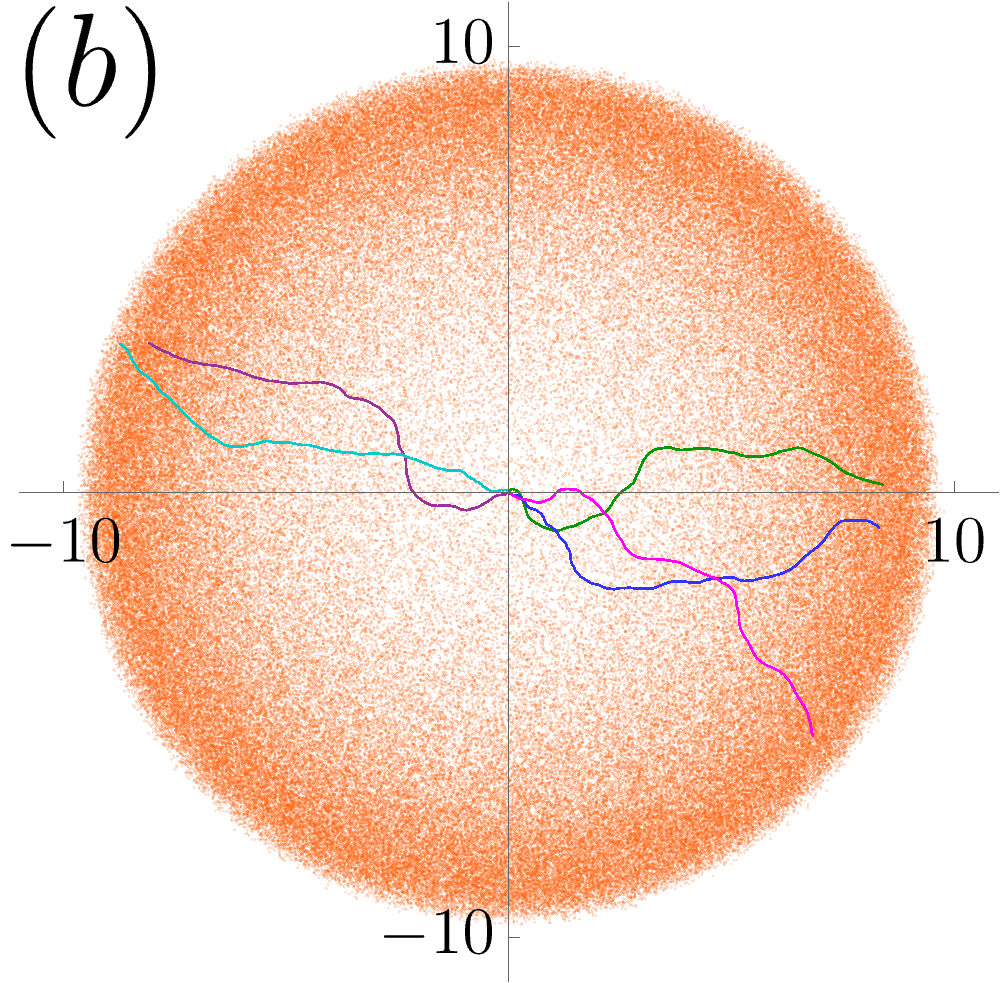}\includegraphics[width=0.2\textwidth]{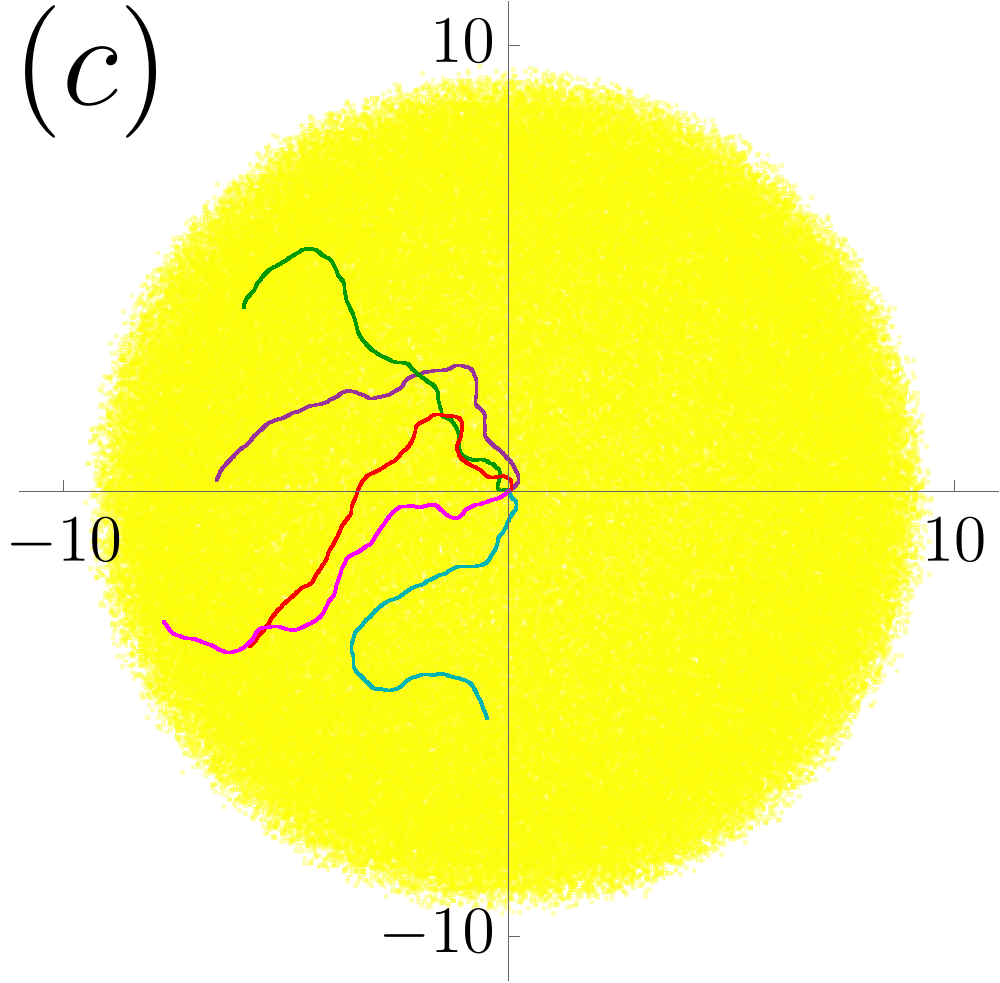}\includegraphics[width=0.2\textwidth]{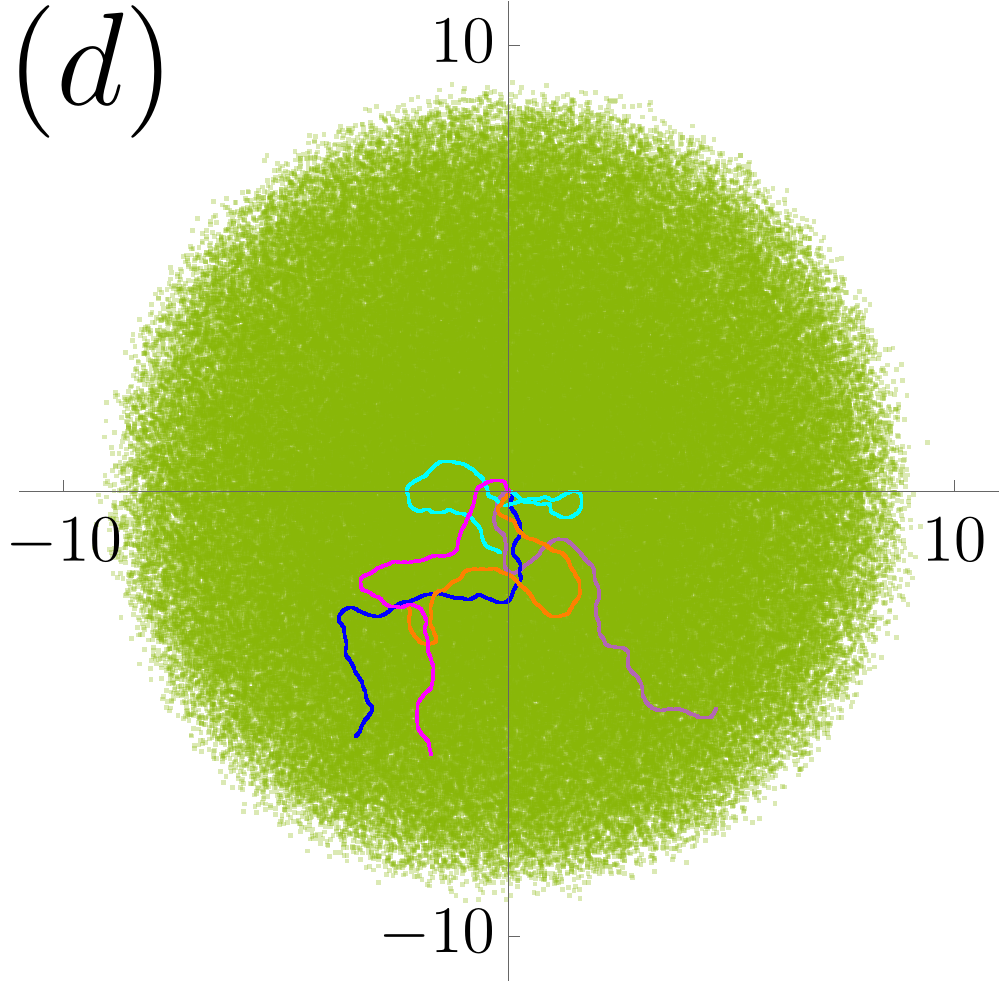}\includegraphics[width=0.2\textwidth]{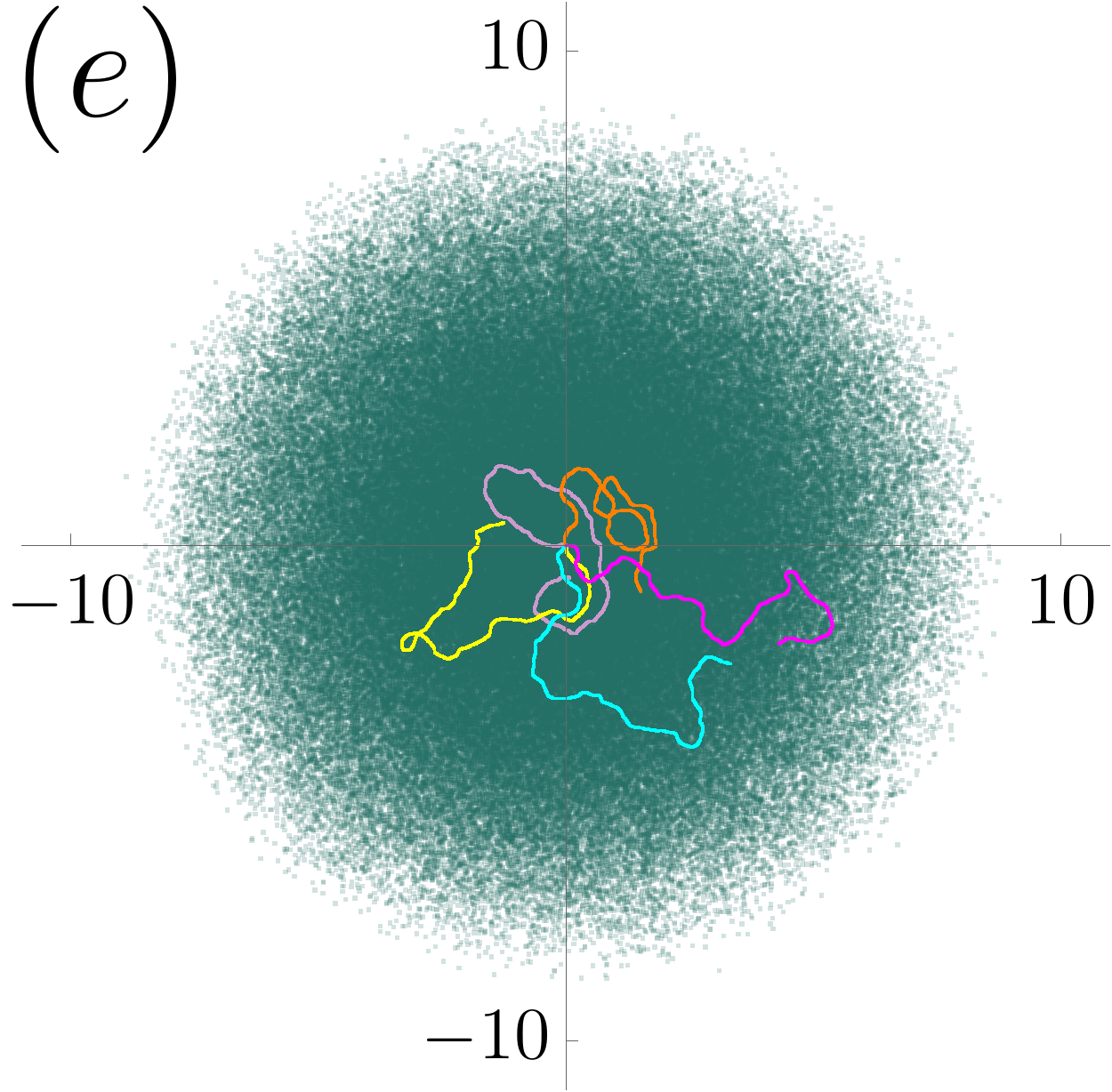}\\
\includegraphics[width=0.2\textwidth]{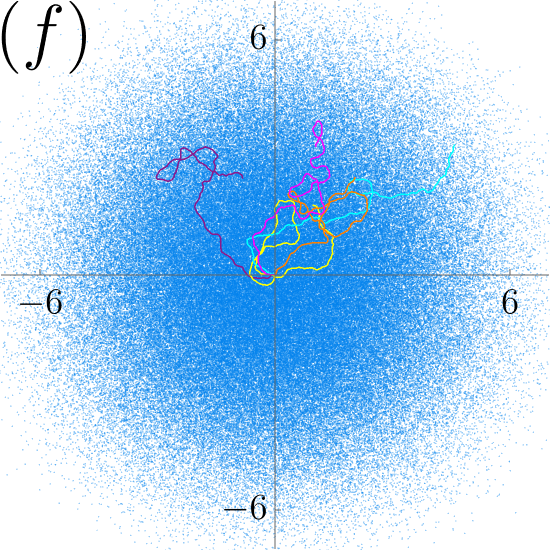}\includegraphics[width=0.2\textwidth]{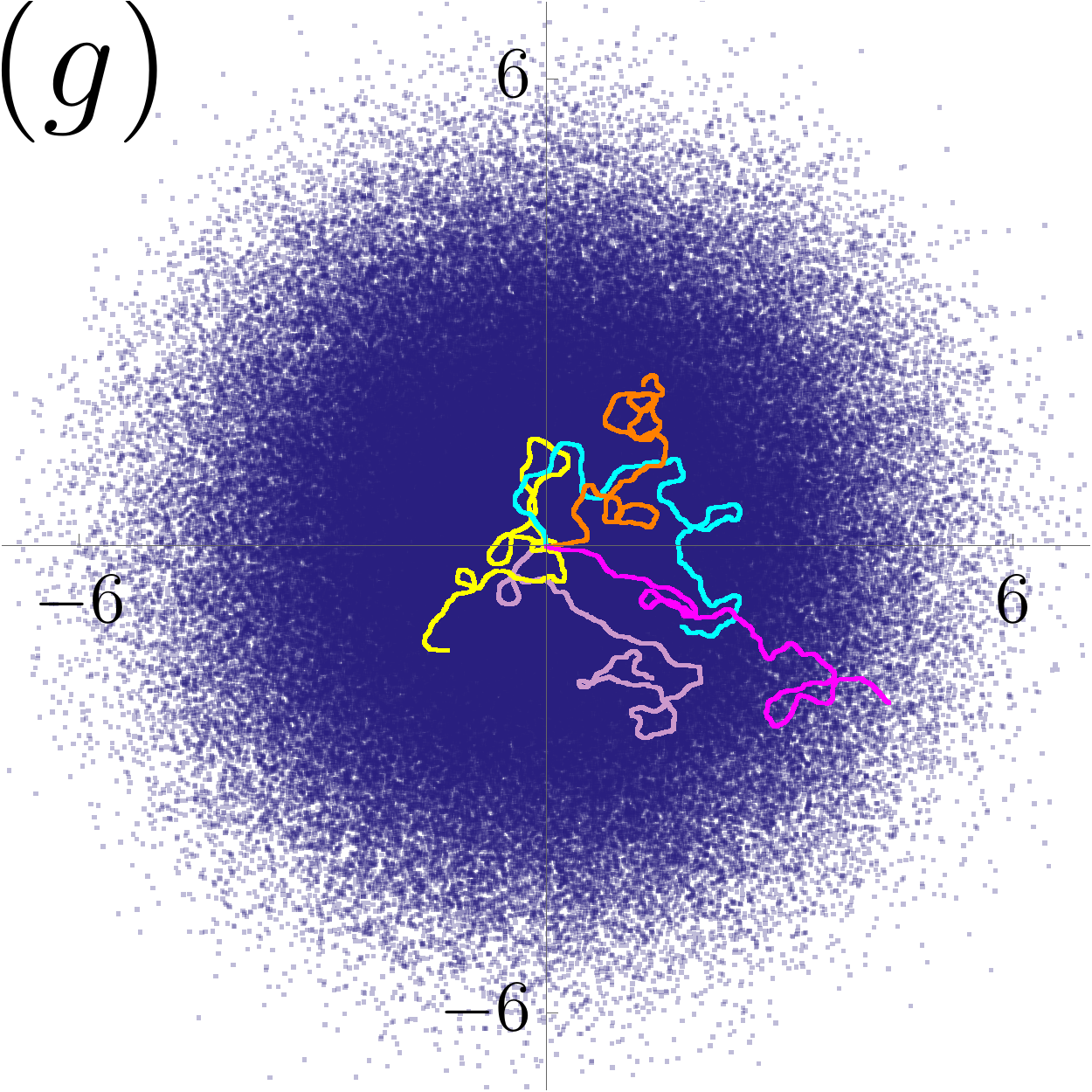}\includegraphics[width=0.2\textwidth]{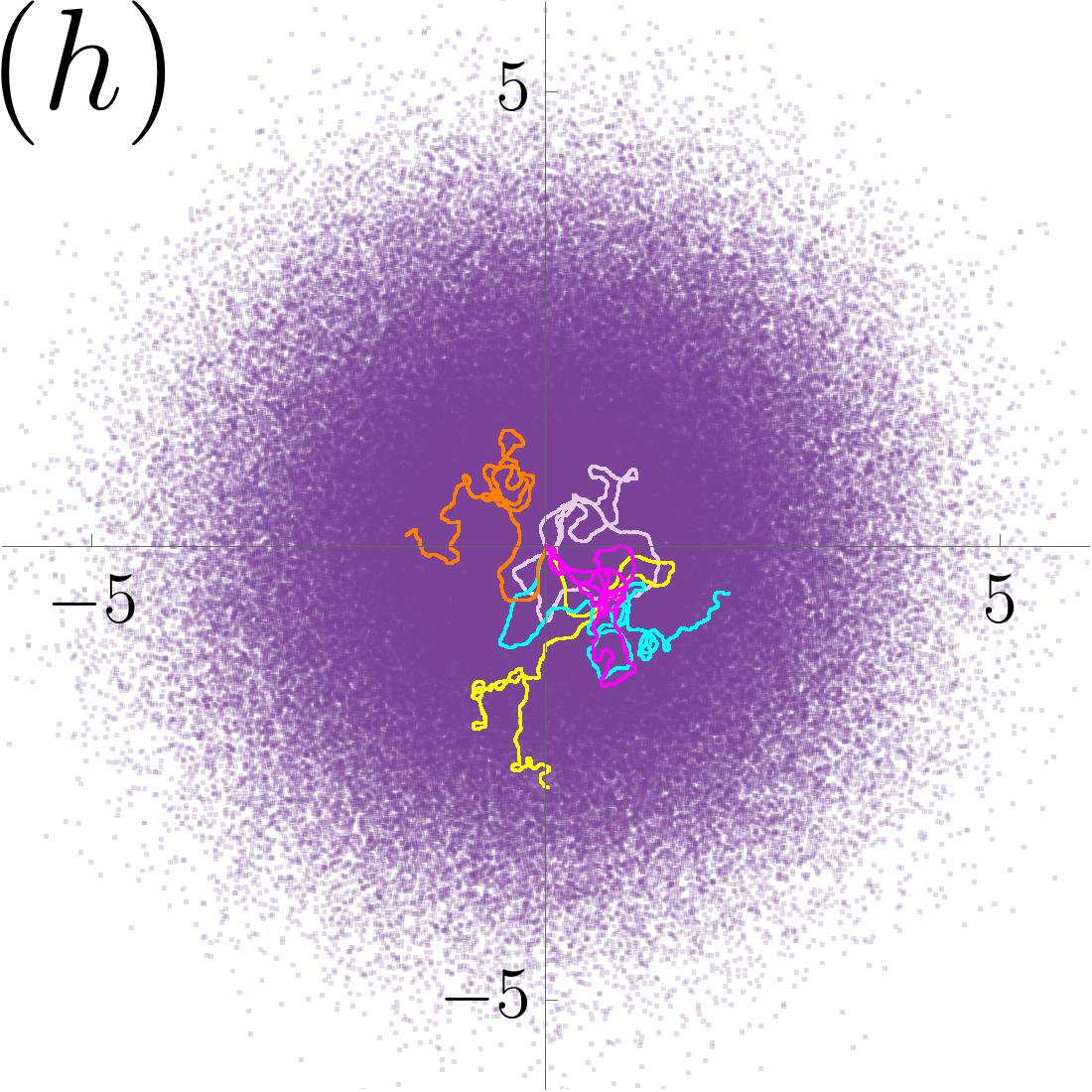}\includegraphics[width=0.2\textwidth]{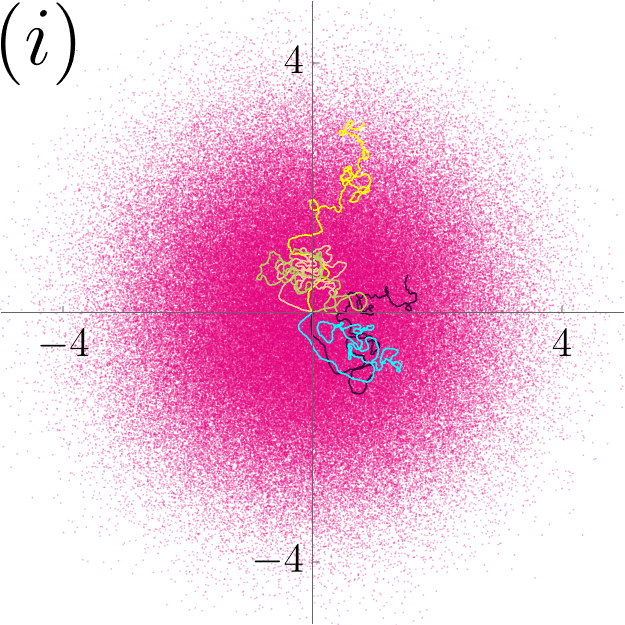}
\caption{Snapshots of the positions of an ensemble of $2.3\times10^{5}$ particles at time $t=10\, R_\beta^{-1/\beta}$ are shown for $\beta=0.2$ ($a$), 0.4 ($b$), 0.6 ($c$), 0.8 ($d$), 1.0 ($e$), 1.2 ($f$), 1.4 ($g$), 1.6 ($h$) and 1.8 ($i$). In each case, five typical trajectories are traced during the time interval $[0,10R_\beta^{1/\beta}]$ all of them starting at the origin of coordinates and initial directions chosen uniformly and random in $[0,\pi]$.}
\label{fig:EnsemblePositions}
\end{figure*}

The associated Fokker-Planck equation for the probability density, $P(\chi,t)$, of finding the outcome $\chi$ at time $t$ for $\chi_{\beta}(t)$,  is given by the Batchelor's equation \cite{Batchelor1952,PostnikovPhysicaA2012,ThielPhysRevE2014b}
\begin{equation}
\frac{\partial}{\partial t}P_\beta(\chi,t)=D_{K,\beta}(t)\frac{\partial^2}{\partial\chi^2}P_\beta(\chi,t)    
\end{equation}
where $D_{K,\beta}(t)=\beta K_\beta t^{\beta-1}$ is the time dependent diffusion coefficient. SBm has been used to describe anomalous diffusion, for instance in the anomalous diffusion properties of fluorescently tagged gold beads in the cytoplasm and the nucleus of living cells \cite{GuigasBioPhysJ2007}, the self-diffusion in granular gases \cite{BrilliantovPhysRevE2000}, in numerical simulations of fluorescence photobleaching recovery experiments \cite{SaxtonBioPhysJ2001} or as a model to fit parameters in fluorescence correlation spectroscopy \cite{WuBioPhysJ2008}. 
Different studies have shown that the statistical properties of sBm give rise to variety of nontrivial effects \cite{SafdariPhysRevE2017,SposiniNewJPhys2019,DosSantosChaosSolFract2021,BodrovaPhysRevE2019a,BodrovaPhysRevE2019b,StojkoskiPhysRevE2021}, particularly under conditions that exacerbate them, for instance, under conditions of confinement \cite{JeonPhysChemChemPhys2014,BodrovaNewJPhys2015} or when particles move on manifolds \cite{Valdes-GomezPhysRevE2023}. 

In section \ref{SectModel} we detail the mathematical model for the analysis, namely active Brownian particles driven by scaled Brownian motion. We give the model as stochastic differential equations from which ensembles of trajectories are obtained by their numerical solution. We also provide the corresponding Fokker-Plank equation for the probability density of finding a particle at position $\boldsymbol{x}$ and propelled in the direction $\hat{v}$ at time $t$, from which analytical results are obtained. In sect \ref{ReducedPDF} we focus our analysis on the reduced probability density which is independent of the propelling direction, from which approximated results for the intermediate scattering function, and kurtosis are obtained. Remarkably the mean squared displacement and propulsion auto-correlation function are obtained in exact manner. We conclude in section \ref{Conclusions}.

%%%%%%%%%%%%%%%%%%%%%%%%%%%%%%%%%%%%%%%%%%%%%%%%%%%%%%%%%%%%%%%%%%%%%%%%%%%%%%%%%%%%%%%%%%%%%%%%%%%%%%%%%%%%%%%%%%%%%%%%%%%%%%%%%%%%%%%%%%%%%
\section{\label{SectModel}The model}

We consider a particle that are propelled in a two-dimensional
domain with constant propulsion speed $v_{0}$ along the direction $\hat{\boldsymbol{v}}(t)=\bigl(\cos \varphi_\beta (t),\sin \varphi_\beta
(t)\bigr)$, $\varphi_\beta (t)$ being the angle between the direction of motion and
the horizontal axis. 
Active fluctuations affects the rotational dynamics leading to rotational diffusion of the propelling ``force'', which in this study is modeled by sBm. The overdamped dynamics of an active Brownian particle follows from the following Langevin equations \cite{tenHagenJPhysCondMatt2011,SevillaPhysRevE2014}
\begin{subequations}
\label{modelo}
\begin{align}
\frac{d{}}{dt}\boldsymbol{x}(t)& =v_0\,\hat{\boldsymbol{v}}(t)%+\boldsymbol{\xi}_{T}(t)
,  \label{LangevinPosition} \\
\frac{d}{dt}\varphi_{\beta}(t)&=\sqrt{2D_{R,\beta}(t)}\,\xi(t),    
\label{LangevinDirection}
\end{align}
\end{subequations}
where $\boldsymbol{x}(t)=\bigl(x_1(t),x_2(t)\bigr)$ denotes the particle position and with $0<\beta$. $\xi(t)$ in Eq. \eqref{LangevinDirection} is Gaussian white noise, $\langle \xi\rangle =0$, $\langle \xi(t)\xi(s)\rangle =\delta(t-s)$ and $D_{R,\beta}(t)=\beta R_\beta t^{\beta-1}$ with $R_\beta$ a constant with units of 1/[Time]$^{\beta}$ gives a measure of the fluctuations at a given time. Standard rotational Brownian motion is recovered in the case $\beta=1$. We introduce the length $l=v_0/R_\beta^{1/\beta}$ as the distance traveled by a particle during the time $R_\beta^{-1/\beta}$. 

The Fokker-Planck equation for the probability
density $P({\boldsymbol{x}},\varphi ,t)\equiv \bigl\langle \delta \bigl({\boldsymbol{x}%
}-{\boldsymbol{x}}(t)\bigr)\delta \bigl(\varphi -\varphi_\beta (t)\bigr)\bigr\rangle$ corresponding to Eqs. \eqref{modelo} is given by
\begin{multline}\label{ActiveFPE}
\frac{\partial }{\partial t}p_\beta({\boldsymbol{x}},\varphi ,t)+v_{0}\nabla\cdot\,\hat{
\boldsymbol{v}}\, p_\beta({\boldsymbol{x}},\varphi,t)= \\ D_{R,\beta}(t)\frac{\partial^{2}}{\partial\varphi^{2}}p_\beta({\boldsymbol{x}}%
,\varphi ,t),
\end{multline}
which correspond to active Brownian motion driven by sBm dynamics. 

%%%%%%%%%%%%%%%%%%%%%%%%%%%%%%%%%%%%%%%%%%%%%%%%%%%%%%%
\section{\label{ReducedPDF}The reduced probability distribution $p_\beta(\boldsymbol{x},t)$}

We now focus our analysis on the reduced distribution of the particle positions (for the active part of motion only as commented in the last section), i.e., we focus on the probability density function of finding a particle at position $\boldsymbol{x}$ at tim $t$ independently of the direction of motion $\varphi$. 

We solved numerically Eqs. \eqref{modelo}, for an ensemble of $N=2.3\times10^5$ particles. The solutions were carried out by use of standard Brownian dynamics up to times $t=10^4\, R_\beta^{-1/\beta}$ using a discretization time step $\Delta t=10^{-3}\, R_\beta^{-1/\beta}$. The positions  of the particles in the ensemble are shown in Fig.~\ref{fig:EnsemblePositions} at time $t=10\, R_\beta^{-1/\beta}$, additionally, four typical trajectories are shown to exhibit the stochastic dynamics for each
$\beta=0.2$, 0.4, 0.6, 0.8, 1.0, 1.2, 1.4 , 1.6 and 1.8. Five trajectories of the ensemble are shown in each case that indicates the corresponding stochastic dynamics. 

First we resort to the Fourier transform of the spatial variables of Eq.~\eqref{ActiveFPE}, i.e. $\boldsymbol{x}=(x_1,x_2)\longrightarrow\boldsymbol{k}=(k_1,k_2)$ 
\begin{multline}\label{ActiveFPE-Fourier}
\frac{\partial}{\partial t}\hat{p}_\beta({\boldsymbol{k}},\varphi
,t)+iv_{0}\,\hat{\boldsymbol{v}}\cdot {\boldsymbol{k}}\, \hat{p}_\beta({
\boldsymbol{k}},\varphi ,t) =\\
D_{R,\beta}(t)\frac{\partial ^{2}}{\partial \varphi ^{2}}\hat{p}({\boldsymbol{k}},\varphi ,t),
\end{multline}
where $\hat{f}({\boldsymbol{k}})=\int d^{2}x\,e^{-i\boldsymbol{k}\cdot \boldsymbol{x}}\,f(\boldsymbol{x})$
denotes the Fourier transform of $f(\boldsymbol{x})$ 
and ${\boldsymbol{k}}=(k_{x},k_{y})$ denotes the Fourier wave-vector. We consider the series expansion
\begin{equation}\label{SeriesExpansion}
\hat{p}_\beta(\boldsymbol{k},\varphi ,t)=\frac{1}{2\pi}\sum\limits_{n=-\infty}
^{\infty }\hat{q}_\beta^{(n)}({\boldsymbol{k}}
,t)\, e^{-R_\beta t^\beta n^{2}}e^{in\varphi},
\end{equation}
where $\bigl\{e^{-R_\beta t^\beta n^{2}}e^{in\varphi}\bigr\}$ with $n$ an integer, is the set of independent solutions of the equation
\begin{equation}\label{Phi}
\frac{\partial}{\partial t}\Phi_\beta(\varphi
,t)=\\
D_{R,\beta}(t)\frac{\partial ^{2}}{\partial \varphi ^{2}}\Phi_\beta(\varphi ,t).
\end{equation}

\begin{figure*}
\includegraphics[width=0.33\textwidth,trim=0 0 0 0,clip]{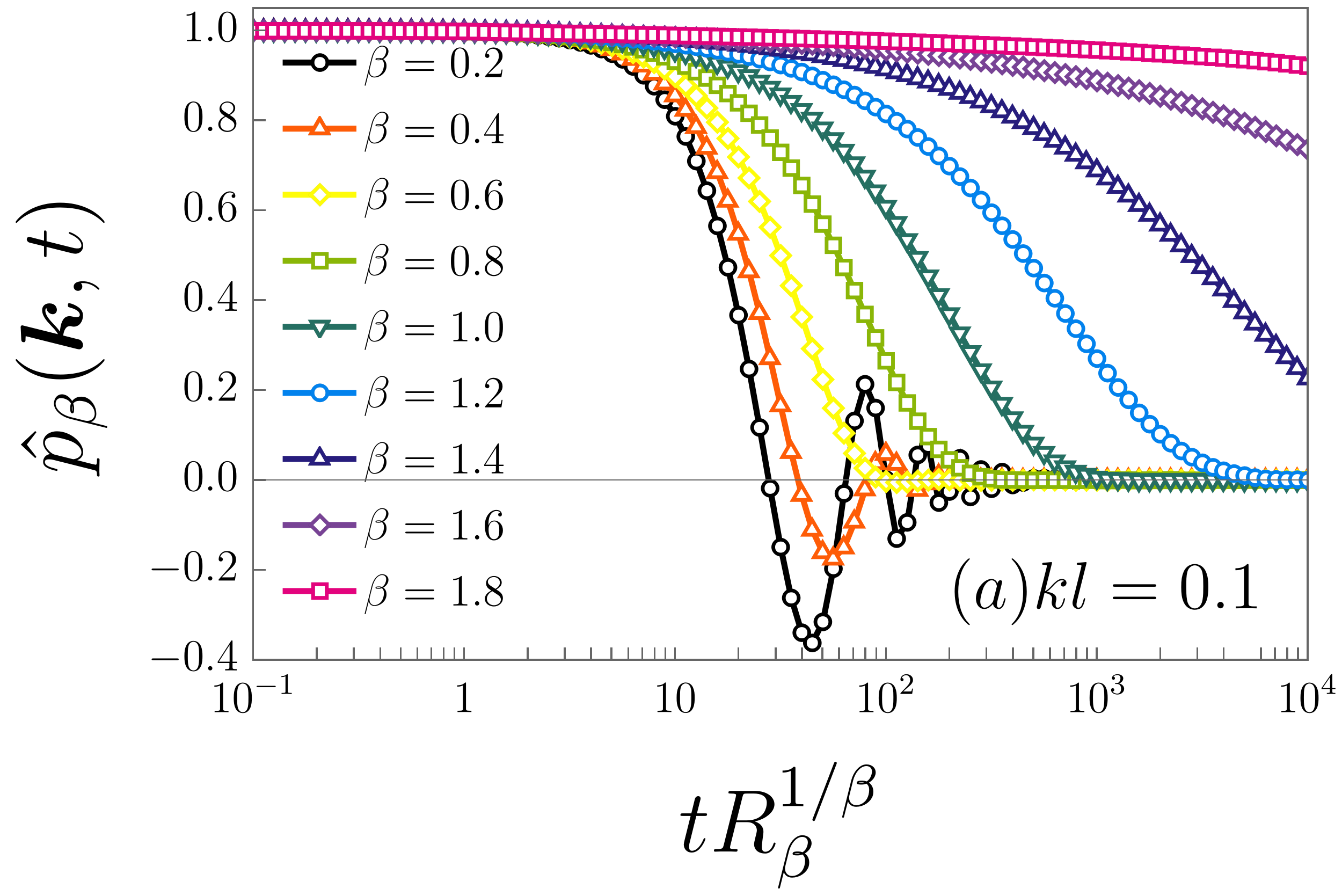}\includegraphics[width=0.33\textwidth,trim=0 0 0 0,clip]{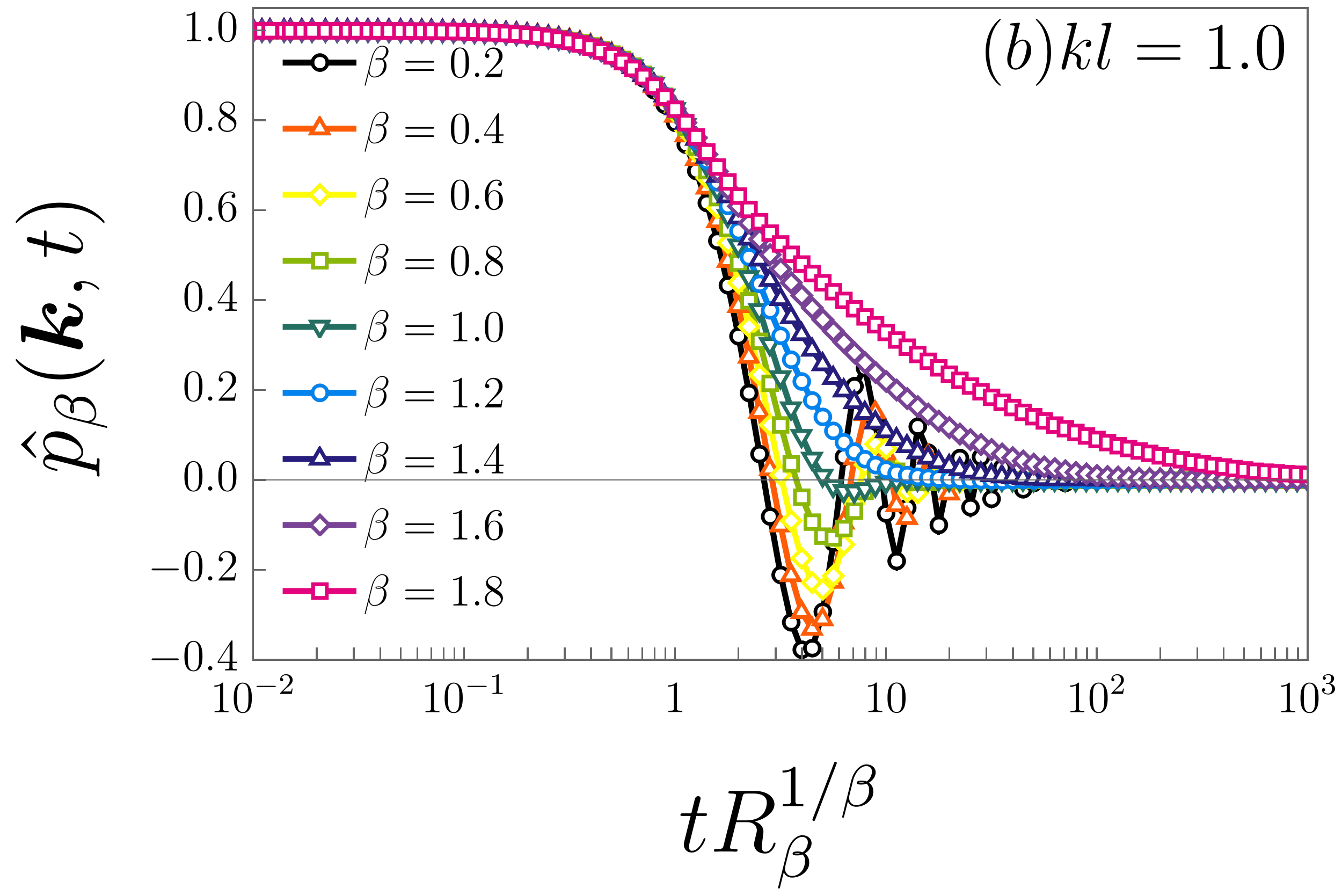}\includegraphics[width=0.33\textwidth,trim=0 0 0 0,clip]{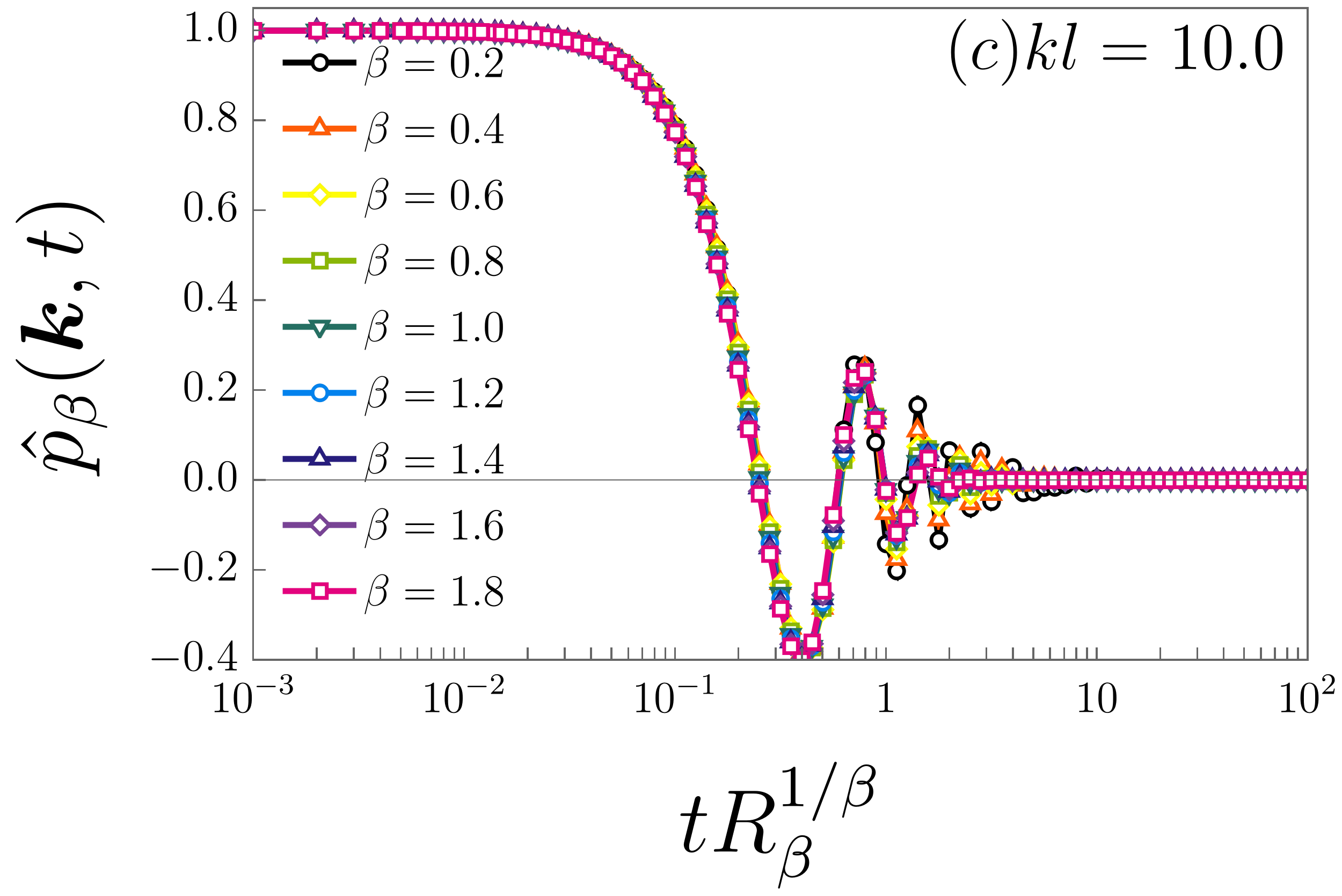}\\
\includegraphics[width=0.33\textwidth,trim=0 0 0 0,clip]{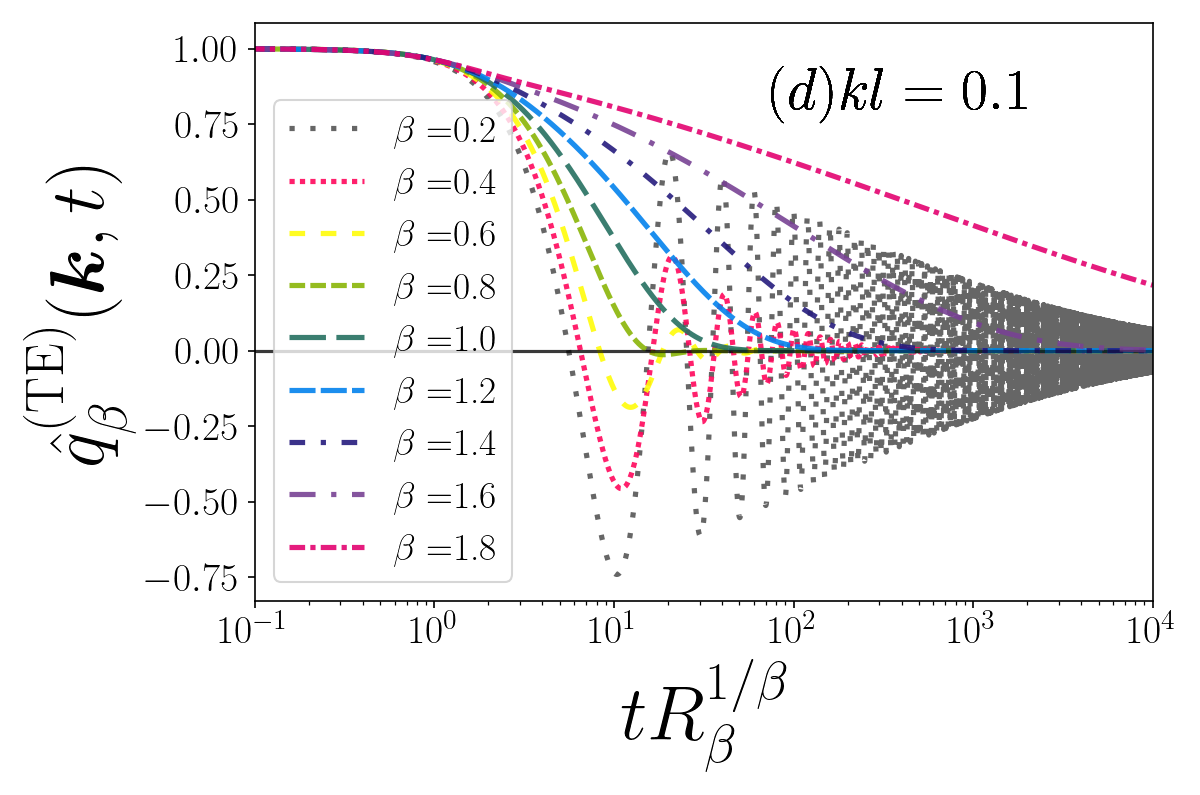}\includegraphics[width=0.33\textwidth,trim=0 0 0 0,clip]{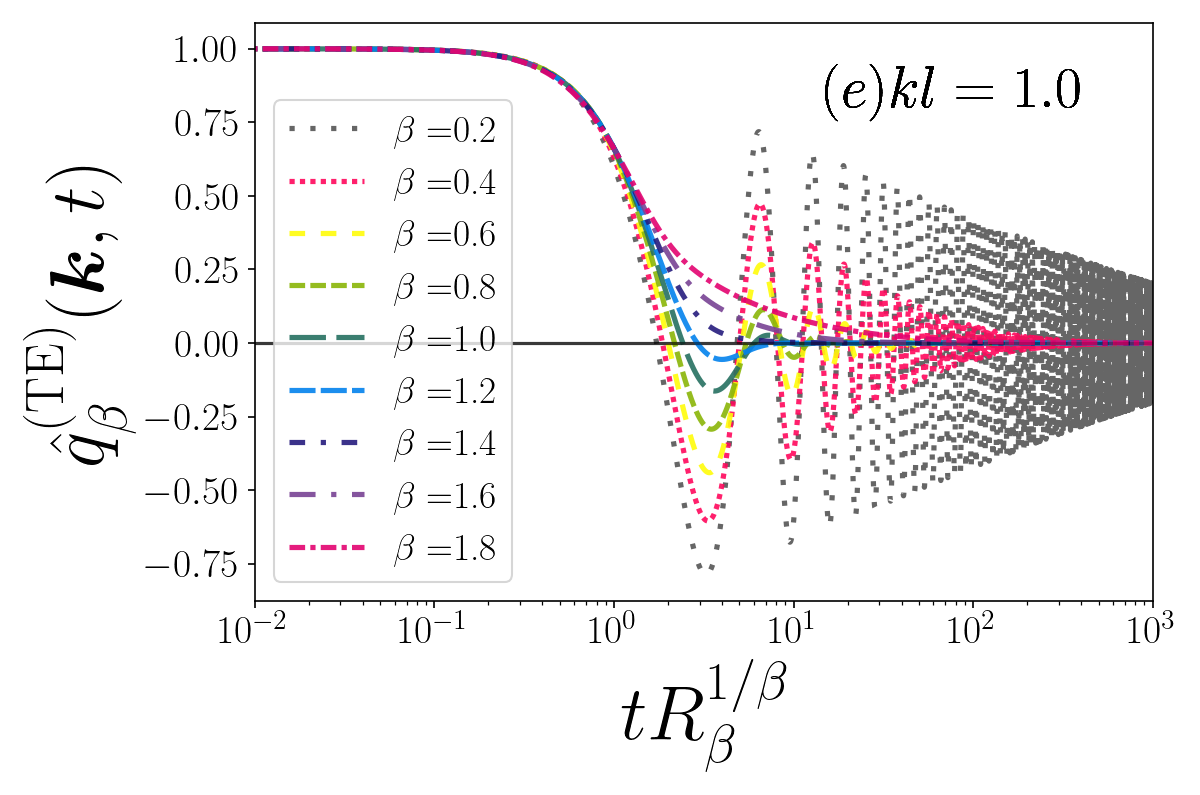}\includegraphics[width=0.33\textwidth,trim=0 0 0 0,clip]{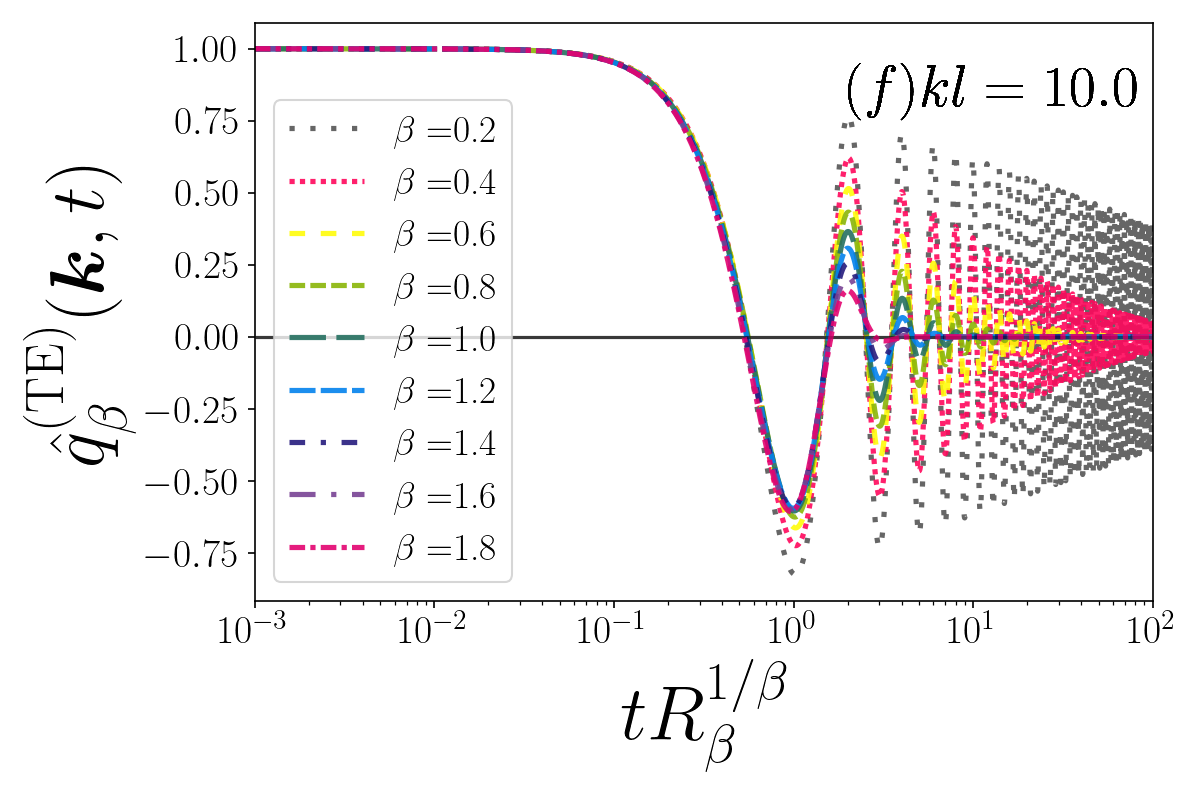}
\caption{The time dependence of the intermediate scattering function. Top row, obtained from numerical simulations, $\hat{p}_\beta(\boldsymbol{k},t)$, for dimensionless wavevectors $kl=0.1$ (a), 1.0 (b) and 10.0 (c). Bottom row, from the numerical integration of Eq.~\eqref{SMoment-Trans}, $\hat{q}_\beta^{(\text{TE})}(\boldsymbol{k},t)$, for dimensionless wavevectors $kl=0.1$ (d), 1.0 (e) and 10.0 (f). }
\label{fig:ISF}
\end{figure*}

The coefficients of the expansion \eqref{SeriesExpansion} are obtained by the use of the standard orthogonality relation
among the Fourier basis functions $\left\{e^{in\varphi}\right\}$, explicitly
\begin{equation}\label{InvCoefficients}
\hat{q}_\beta^{(n)}({\boldsymbol{k}},t)=e^{R_\beta t^\beta n^{2}}\int_{-\pi}^{\pi}d\varphi
\, \hat{p}_\beta({\boldsymbol{k}},\varphi ,t)e^{-in\varphi }.
\end{equation}
From this we have 
\begin{equation}\label{q0}
\hat{q}_\beta^{(0)}(\boldsymbol{k},t)=\int_{-\pi}^\pi d\varphi\,\hat{p}_\beta(\boldsymbol{k},\varphi,t)=\hat{p}_\beta(\boldsymbol{k},t)
\end{equation}
which corresponds to the self or coherent part of the Intermediate Scattering Function (ISF). This is obtained from numerical simulations and shown in Fig.~\ref{fig:ISF} for small  wave-vector number $kl=0.1$ (a), intermediate $kl=1.0$ 
(b) and large wave-vector number $kl=10.0$ (c), in all cases, the ballistic short time behavior ($\hat{p}_{\beta} \approx 1$) is clearly observed. 

In the case $kl=10$, ISF characterizes the particles position distribution at short-length scales, at short times it relaxes in the same manner independently of the value of $\beta$, in this regime particle distributions are all alike as the one shown in Fig.~\ref{fig:EnsemblePositions}(a), corresponding to ballistic propagation. After times of the order of $tR_\beta^{1/\beta}\gtrsim 1$, relaxation of the ISF do depends on $\beta$ as shown by the long-lasting oscillations the smaller $\beta$ is. 
For the cases for which the wave-vector number is of the order or smaller than  $l^{-1}$, $kl\lesssim1$ ($kl=1.0$ and $kl=0.1$ ar shown in Fig.~\ref{fig:ISF}(b) and (a), respectively), it is possible to identify the effects of the different values of $\beta$ in the intermediate-time regimes, i.e., the ISF relaxation distinguishes among the distinct patterns of propagation induced by the different sBm exponent $\beta$. For $kl=1.0$ in Fig. \ref{fig:ISF} (b) the ISF oscillates decaying to zero faster if $\beta\lesssim 1.0$, instead it decays slowly and monotonically when $\beta>1$, meaning that the corresponding relaxation time increases. These features are more noticeable the smaller the values of $kl$ are. In Fig.~\ref{fig:ISF} (a) the case $kl=0.1$ is shown, now the  oscillating decay of the ISF is observed for $\beta\lesssim0.6$ and monotonically decaying ISF's  for $\beta>0.6$.
Similar characteristics have been reported in the case $\beta=1$ for standard active Brownian motion \cite{Kurzthaler2006,Kurzthaler2017}. Fig. \ref{fig:ISF}(d)-(f) also includes approximate results for the ISF  computed from a generalization of the telegrapher equation, which is discussed in the next section.

The moments of $p_\beta(\boldsymbol{x},t)$ are of interest and can be obtained from its characteristic function $\hat{p}_\beta(\boldsymbol{k},t)=\hat{q}_\beta^{(0)}(\boldsymbol{k},t)$ as 
\begin{multline}\label{moments}
\bigl\langle x_1^{m_1}x_2^{m_2}\bigr\rangle=\Biggl[\biggl(-i\frac{\partial}{\partial k_1}\biggr)^{m_1}\times\\
\biggl(-i\frac{\partial}{\partial k_2}\biggr)^{m_2}\hat{q}_\beta^{(0)}(\boldsymbol{k},t)\Biggr]_{\substack{k_1=0\\k_2=0}}.
\end{multline}

After substitution of Eq.~\eqref{SeriesExpansion} into Eq.~\eqref{ActiveFPE-Fourier} and after use of the orthogonality of the
Fourier basis functions we get the following set of coupled ordinary differential
equations for the $n$-th coefficient of the expansion $\hat{q}_\beta^{(n)}({
\boldsymbol{k}},t),$ namely
\begin{multline}\label{Hierarchy}
\frac{d}{dt}\hat{q}_\beta^{(n)}(\boldsymbol{k};t) =-i\frac{v_{0}}{2}k\Bigl[e^{-i\theta}
e^{-R_\beta t^\beta(1-2n)}\,\hat{q}_\beta^{(n-1)}(\boldsymbol{k};t)
\\
 +e^{i\theta}e^{-R_\beta t^\beta(1+2n)}\,\hat{q}
_\beta^{(n+1)}(\boldsymbol{k};t)\Bigr],
\end{multline}
$\theta$ and $k=\vert\boldsymbol{k}\vert$ are defined through the relations $k_{1}\pm ik_{2}=ke^{\pm i\theta}$.

\subsection{The $\hat{q}_\beta^{(1)}$ approximation: The telegrapher equation with time-dependent relaxation coefficient}

We can close the coupled Eqs. \eqref{Hierarchy} up to $\hat{q}_\beta^{(\pm1)}(\boldsymbol{k};t)$, disregarding, as first approximation, all coefficients $\hat{q}_\beta^{(n)}(\boldsymbol{k};t)$ with $\vert n\vert>1$ \cite{HeizlerNucSciEngin2010,SevillaPRE2015}. This approximation is expected to be appropriate in the long-time regime since high Fourier modes in the expansion \eqref{SeriesExpansion} are negligible small in that limit, however for $0<\beta<\frac{1}{2}$ is not the case due to the slow dynamics. Notwithstanding this, the exact time dependence of the MSD is obtained from this as is shown afterwards. We have that in this approximation 
\begin{subequations}\label{q1approx}
\begin{align}
\frac{\partial}{\partial t}\hat{q}_\beta^{(1)}(\boldsymbol{k};t)&=-i\frac{v_0}{2}ke^{-i\theta}e^{R_{\beta}t^\beta}\hat{q}_\beta^{(0)}(\boldsymbol{k};t),\\
\frac{\partial}{\partial t}\hat{q}_\beta^{(0)}(\boldsymbol{k};t)&=-i\frac{v_0}{2}ke^{-R_{\beta}t^\beta}\bigl[e^{-i\theta}\hat{q}_\beta^{(-1)}(\boldsymbol{k};t)+e^{i\theta}\hat{q}_\beta^{(1)}(\boldsymbol{k};t)\bigr],\\
\frac{\partial}{\partial t}\hat{q}_\beta^{(-1)}(\boldsymbol{k};t)&=-i\frac{v_0}{2}ke^{i\theta}e^{R_{\beta}t^\beta}\hat{q}_\beta^{(0)}(\boldsymbol{k};t).
\end{align}    
\end{subequations}
The $\hat{q}_\beta^{(0)}(\boldsymbol{k};t)$ in Eq.~\eqref{q0} must be distinguished from the corresponding one solution of Eqs.~\eqref{q1approx}, which is an approximation for the former and we refer to it as $\hat{q}_\beta^{(\text{TE})}(\boldsymbol{k};t)$. This is given by
\begin{equation}\label{GDE-Fourier}
\frac{\partial}{\partial t}\hat{q}_\beta^{(\text{TE})}(\boldsymbol{k};t)=-\frac{v_0^2}{2}k^2e^{-R_{\beta}t^\beta}\int_0^tds\, e^{R_\beta s^\beta}\hat{q}_\beta^{(\text{TE})}(\boldsymbol{k};s),  
\end{equation}
with $k^2=k_1^2+k_2^2$. After inverting the Fourier transform we obtain the \emph{generalized diffusion equation} 
\begin{equation}
\frac{\partial}{\partial t}q_\beta^{(\text{TE})}(\boldsymbol{x},t)=\frac{v_0^2}{2}\int_{0}^t ds\,\gamma(t;s)\nabla^2 q_\beta^{(\text{TE})}(\boldsymbol{x},s),
\end{equation}
where the memory function $\gamma(t;s)=e^{-R_\beta(t^\beta-s^\beta)}$ is not time-translational invariant, a characteristic that emerges from the inherently nonstationarity of the sBm process.

After taking the derivative with respect time we get, equivalently 
\begin{equation}\label{SMoment-Trans}
 \frac{\partial^{2}}{\partial t^{2}}\hat{q}_\beta^{(\text{TE})}(\boldsymbol{k};t)+D_{R,\beta}(t)\frac{\partial}{\partial t}\hat{q}_\beta^{(\text{TE})}(\boldsymbol{k};t)=-\frac{v_{0}^{2}}{2}k^2\hat{q}_\beta^{(\text{TE})}(\boldsymbol{k};t),
\end{equation}
which corresponds to a damped harmonic oscillator with time dependent damping coefficient $D_{R,\beta}(t)$,
whose inverse Fourier transform is identified with the telegrapher's equation with time dependent relaxation coefficient $D_{R,\beta}(t)$
\begin{equation}
 \frac{\partial^{2}}{\partial t^{2}}q_\beta^{(\text{TE})}(\boldsymbol{x};t)+D_{R,\beta}(t)\frac{\partial}{\partial t}q_\beta^{(\text{TE})}(\boldsymbol{x};t)=\frac{v_{0}^{2}}{2}\nabla^{2}q_\beta^{(\text{TE})}(\boldsymbol{x};t).
\end{equation}
Differential equations that determine the time dependence of the moments of $q_\beta^{(\text{TE})}(\boldsymbol{x},t)$, can be obtained by using Eq.~\eqref{GDE-Fourier}, but with $q_\beta^{(\text{TE})}(\boldsymbol{x},t)$ instead of $q_\beta^{(0)}(\boldsymbol{x},t)$. 

In panels (d)-(f) of Fig. \ref{fig:ISF} the time dependence of the ISF, obtained from the numerical solution of the telegrapher equation with time-dependent relaxation coefficient \eqref{SMoment-Trans}, is shown for different values of the dimensionless wavevector $kl$ and different scaling parameter $\beta$. Results are qualitatively in good agreement with those discussed in the previous section. The main differences with respect to the exactly computed cases (Fig.~\ref{fig:ISF} (a)-(c)) are that relaxation times increase orders of magnitude and oscillation amplitudes become larger for $\beta$ ($\lesssim0.5$); and when the approximated ISF decays monotonically, the decaying is faster than in the exact case. This analysis elucidates the importance of the higher Fourier modes left aside in the approximation \eqref{q1approx}. Notwithstanding this, the approximation made in \eqref{q1approx} leads to the exact time dependence of the MSD as discussed in the following section.

\subsection{The mean squared displacement}
\begin{figure}[t]
 \includegraphics[width=\columnwidth]{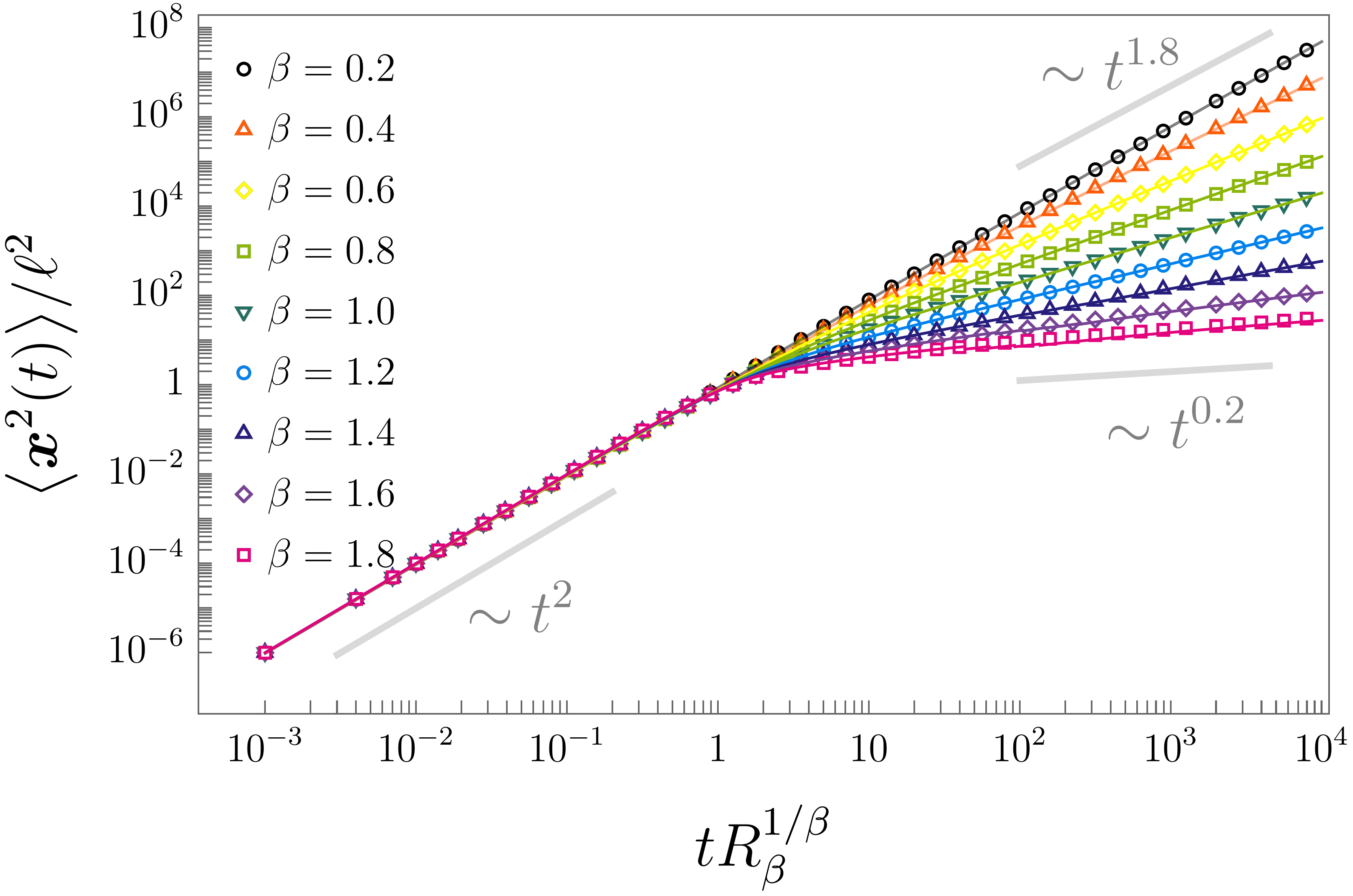}
 \caption{Time dependence of the dimensionless mean squared displacement $\bigl\langle\boldsymbol{x}^2(t)\bigr\rangle/\ell^2$, $\ell=v_0/R_{\beta}^{1/\beta}$ being the \emph{persistence length}. Symbols show the exact result $\bigl\langle\boldsymbol{x}^2(t)\bigr\rangle$ from our numerical analysis, while continuous lines mark the analytical expression $\bigl\langle\boldsymbol{x}^2(t)\bigr\rangle_\text{TE}$ \eqref{MSD-Analytical} obtained from the $\hat{q}_\beta^{(1)}$ approximation. The agreement is remarkable.}
 \label{Fig1_msd}
\end{figure}

The diffusive transport of the particle is generically characterized by the second moment or mean-squared displacement of the position distributions $p_\beta(\boldsymbol{x},t)$, $\langle\boldsymbol{x}^2(t)\rangle=\int d^2x\, \boldsymbol{x}^2p_\beta(\boldsymbol{x},t)$ and accordingly to Eqs.~\eqref{moments} and \eqref{q0}, is given by
\begin{align}
\bigl\langle\boldsymbol{x}^2(t)\bigr\rangle_{p_\beta}=&
\bigl\langle x_1^2(t)\bigr\rangle_{p_\beta}+\bigl\langle x_2^2(t)\bigr\rangle_{p_\beta}\nonumber\\
=&-\Biggl[\biggl(\frac{\partial^2}{\partial k_1^2}+\frac{\partial^2}{\partial k_2^2}\biggr)\hat{q}_\beta^{(0)}(\boldsymbol{k},t)\Biggr]_{\substack{k_1=0\\k_2=0}}.
\end{align}
After applying the differential operator of last equation to the Fourier transform of the generalized diffusion equation~\eqref{GDE-Fourier}, the approximation $\hat{q}_\beta^{(0)}(\boldsymbol{k},t)\approx \hat{q}_\beta^{(\text{TE})}(\boldsymbol{k},t)$ leads to the result  
\begin{equation}\label{MSD-ode}
\frac{d}{dt}\bigl\langle\boldsymbol{x}^{2}(t)\bigr\rangle_\text{TE}=2v_{0}^{2}e^{-R_\beta t^\beta}\int_0^t ds\, e^{R_\beta s^\beta},
\end{equation}
from which we get straightforwardly that
\begin{equation}\label{MSD-Analytical}
 \langle\boldsymbol{x}^{2}(t)\rangle_\text{TE}=2v_{0}^{2}\int_{0}^{t}ds\, e^{-R_\beta s^\beta}\int_{0}^{s}ds_{1}\, e^{R_\beta s_{1}^\beta},
\end{equation}
where we have made explicit the initial condition $\bigl\langle\boldsymbol{x}^{2}(0)\bigr\rangle_\text{TE}=0$. This is one of the main results of our analysis. Expression \eqref{MSD-Analytical} is evaluated numerically for $\beta=0.2$, 0.4, 0.6, 0.8, 1.0, 1.2, 1.4, 1.6 and 1.8 and shown in Fig.~\ref{Fig1_msd} (solid lines) and compared with the exact results obtained from the numerical analysis of the trajectories ensemble obtained from integration of Eqs.~\eqref{modelo} (symbols). The agreement is remarkable corroborating that $\hat{q}_\beta^{(\text{TE})}(\boldsymbol{k},t)$ gives the exact time dependence of the mean-squared displacement.

Although not evident at first a glance, analytical Eq. \eqref{MSD-Analytical} masks a crossover between ballistic transport at the short-time regime $\langle\boldsymbol{x}^2(t)\rangle\approx v_0^2 t^2$, to anomalous diffusion $\langle\boldsymbol{x}^2(t)\rangle\sim t^{2-\beta}$ in the long-time one (see Fig.~\ref{Fig1_msd}). This can be seen from the following heuristic argument. By taking the time derivative to Eq.~\eqref{MSD-ode} we get
\begin{equation}\label{MSD-2ode}
\frac{d^2}{dt^2}\langle\boldsymbol{x}^{2}(t)\rangle_\text{TE}+D_{R,\beta}(t)\frac{d}{dt}\langle\boldsymbol{x}^{2}(t)\rangle_\text{TE}=2v_0^2. 
\end{equation}
In the short-time regime, i.e., to times up to the order of $R_\beta^{1/\beta}$, the effects of noise expressed in the second term of Eq.~\eqref{SMoment-Trans} are unimportant compared with the coherent term expressed by the second-order time derivative, and thus we disregard it from Eq.~\eqref{MSD-2ode} to get approximately 
\begin{equation}
\frac{d^2}{dt^2}\langle\boldsymbol{x}^{2}(t)\rangle_\text{TE}=2v_0^2,
\end{equation}
whose solution with the corresponding initial conditions is $\langle\boldsymbol{x}^{2}(t)\rangle_\text{TE}=v_0^2t^2$. Analogously, in the long-time regime, the coherent motion described by the second order time derivative becomes negligible with respect to the effects of fluctuations, thus we have that approximately 
\begin{equation}
D_{R,\beta}(t)\frac{d}{dt}\langle\boldsymbol{x}^{2}(t)\rangle_\text{TE}=2v_0^2 
\end{equation} 
which gives $\langle\boldsymbol{x}^{2}(t)\rangle_\text{TE}=\frac{2v_0^2}{R_\beta \beta(2-\beta)}t^{2-\beta}$. 

A series expansion in powers of $R_\beta^{1/\beta} t$ can be obtained by use of the series expansion of the stretched exponentials in Eq.~\eqref{MSD-ode}, after carrying out the integrals we get
\begin{equation}
\langle\boldsymbol{x}^2(t)\rangle_\text{TE}=v_0^2 t^2\sum_{n,m=0}^\infty \frac{2(-1)^n \bigl(R_\beta t^{\beta}\bigr)^{n+m}}{n!\,m!(\beta m+1)\bigl(\beta(n+m)+2\bigr)}
\end{equation}
which clearly leads to the ballistic transport when $R_\beta^{1/\beta} t\ll1$. Conversely, an asymptotic expansion of the integral in \eqref{MSD-ode} can be obtained after use of the change of variable $s^{\beta}=t^\beta-R_\beta z$, thus it can be written as
\begin{equation*}
e^{R_\beta t^\beta}\frac{t^{1-\beta}}{R_\beta\beta}\int_0^{R_\beta t^\beta}dz\Big(1-\frac{z}{R_\beta t^{\beta}}\Bigr)^{1/\beta-1}e^{-z}.
\end{equation*}
Therefore, in the limit $R_\beta t^\beta\rightarrow\infty$ we have
\begin{equation}
\int_0^t ds\, e^{R_\beta s^\beta}\sim e^{R_\beta t^\beta}\frac{t^{1-\beta}}{R_\beta\beta}\biggl[1+ O\biggl(\frac{1}{R_\beta t^\beta}\biggr)\biggr].
\end{equation}
By retaining the leading term in the asymptotic expansion, we substitute it in Eq.~\eqref{MSD-ode}, and after integration we obtain 
\begin{equation}
\langle\boldsymbol{x}^2(t)\rangle_\text{TE}\sim\frac{2v_0^2}{R_\beta\beta(2-\beta)}t^{2-\beta}
\end{equation}
as mentioned before.

\subsection{The Kurtosis}
The kurtosis of the bivariate distribution $p_\beta(\boldsymbol{x};t)$ is given by \cite{Mardia74p115}
\begin{equation}
\kappa_\beta(t)=\biggl\langle \Bigl[\bigl(\boldsymbol{x}(t)-\langle \boldsymbol{x}(t)\rangle\bigr)\Sigma^{-1}\bigl(\boldsymbol{x}(t)-\langle 
\boldsymbol{x}(t)\rangle\bigr)^{\text{T}}\Bigr]
^{2}\biggr\rangle,  \label{kurtosis}
\end{equation}
where $\boldsymbol{x}^{\text{T}}$ denotes a column vector, transpose of the row vector $\boldsymbol{x}=(x_1,x_2)$ and $\Sigma $ is the $2\times2$ matrix defined by the average of the dyadic product 
$\bigl(\boldsymbol{x}(t)-\langle \boldsymbol{x}(t)\rangle\bigr)^{\text{T}}\cdot \bigl(\boldsymbol{x}(t) -\langle \boldsymbol{x}(t)\rangle\bigr)$. It can be shown that due to the invariance of $p_\beta(\boldsymbol{x},t)$ under spatial rotations, the kurtosis \eqref{kurtosis} reduces to 
\begin{equation}
\kappa_\beta(t)=\frac{\bigl\langle x_1^4(t)\bigr\rangle}{\bigl\langle x_1^2(t)\bigr\rangle^2}+\frac{\bigl\langle x_2^4(t)\bigr\rangle}{\bigl\langle x_2^2(t)\bigr\rangle^2}+\frac{2\bigl\langle x_1^2(t)\,x_2^2(t)\bigr\rangle}{\bigl\langle x_1^2(t)\bigr\rangle\,\bigl\langle x_2^2(t)\bigr\rangle}.
\end{equation}
Each moment involved is by use of the formula~\eqref{moments}, the moments in the denominators are the squared of the second moments computed in the calculation of the MSD. The exact fourth-order moments are given by
\begin{subequations}\label{4th-moments}
\begin{align}
\bigl\langle x_i^4(t)\bigr\rangle=&\biggl[\frac{\partial^4}{\partial k_i^4}\hat{q}_\beta^{(0)}(\boldsymbol{k};t)\biggr]_{\substack{k_1=0\\k_2=0}},\, (i=1,2),\\
\bigl\langle
x_1^2(t)\,x_2^2(t)\bigr\rangle=&\biggl[\frac{\partial^4}{\partial k_1^2\partial x_2^2}\hat{q}_\beta^{(0)}(\boldsymbol{k};t)\biggr]_{\substack{k_1=0\\k_2=0}}.
\end{align}
\end{subequations}

\begin{figure}
\includegraphics[width=\columnwidth]{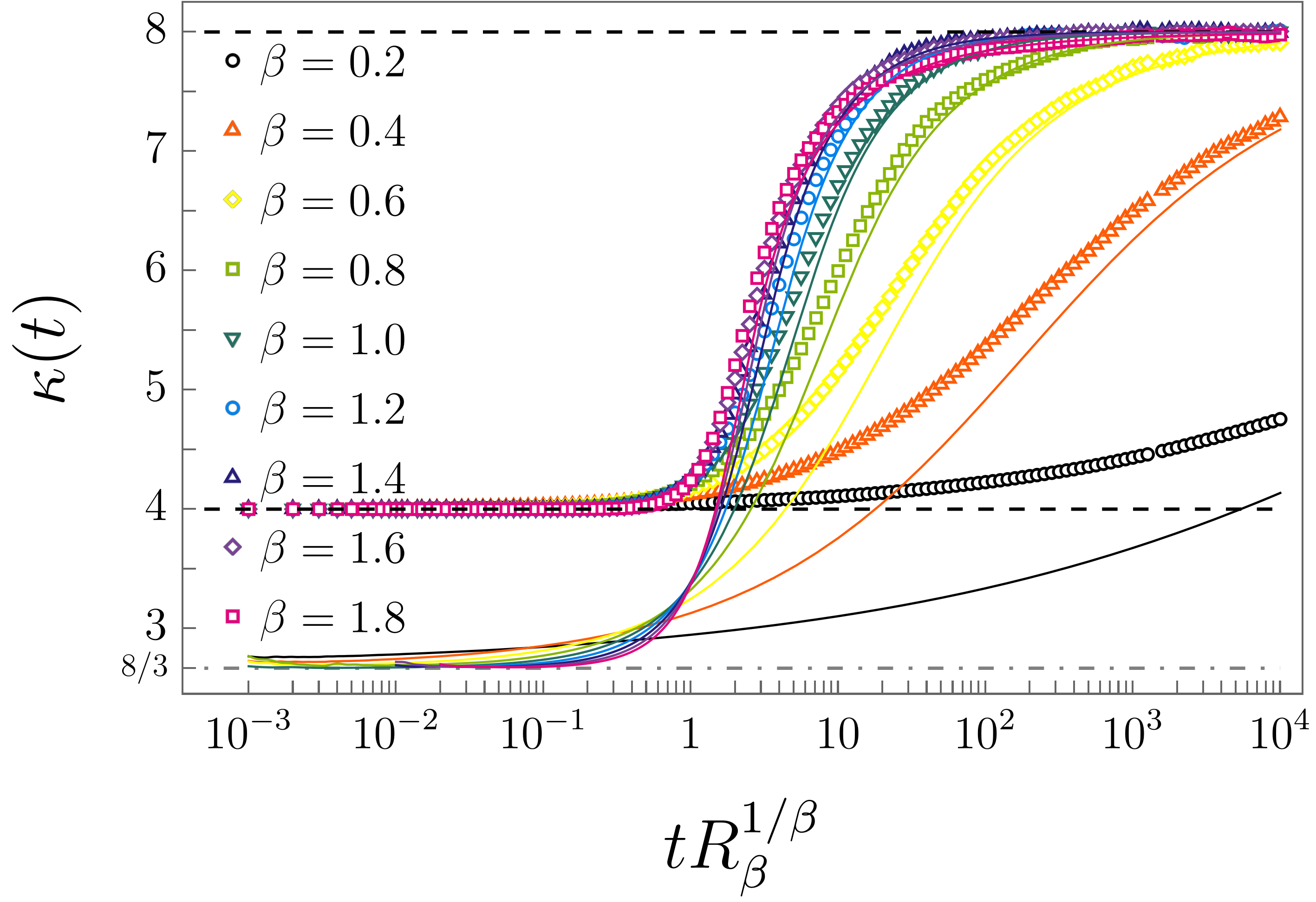}
\caption{Time dependence of the kurtosis of $p_\beta(\boldsymbol{x},t)$ as defined in Eq.~\eqref{kurtosis}. Symbols mark the result from the ensemble average of the trajectories obtained form numerical simulations for $\beta=0.2$, 0.4, 0.6, 0.8, 1.0, 1.2, 1.4, 1.6 and 1.8.}
 \label{fig:kurtosis}
\end{figure}

In Fig.~\eqref{fig:kurtosis} the exact time dependence of $\kappa(t)$, obtained from our numerical analysis,
is shown (symbols) for different values of $\beta$ ranging from 0.2, to 1.8. The short-time regime corresponds to the propagation of a ring of kurtosis $4$ transiting to the value 8, characteristic of a Gaussian in the long-time regime.

Differential equations for the fourth moments in the $\hat{q}_\beta^{(1)}$-approximation can be derived after applying the fourth-order derivatives in Eqs. \eqref{4th-moments} to Eq. \eqref{GDE-Fourier}, we get after some rearrangements
\begin{subequations}
\begin{multline}
\frac{d^2}{dt^2}\bigl\langle x_i^4(t)\bigr\rangle_\text{TE}+D_{R,\beta}(t)\frac{d}{dt}\bigl\langle x_i^4(t)\bigr\rangle_\text{TE}=\\
6v_0^2\bigl\langle x_i^2(t)\bigr\rangle_\text{TE},
\end{multline}
and 
\begin{multline}
\frac{d^2}{dt^2}\bigl\langle x_1^2(t)x_2^2(t)\bigr\rangle_\text{TE}+D_{R,\beta}(t)\frac{d}{dt}\bigl\langle x_1^2(t)x_2^2(t)\bigr\rangle_\text{TE}=\\
v_0^2\Bigl[\bigl\langle x_1^2(t)\bigr\rangle_\text{TE}+\bigl\langle x_2^2(t)\bigr\rangle_\text{TE}\Bigr], 
\end{multline}
\end{subequations}
whose solutions lead to the time dependence of the kurtosis in the $\hat{q}_\beta^{(1)}$-approximation 
\begin{widetext}
\begin{equation}
\kappa_{\beta,\text{TE}}(t)=8
\frac{\displaystyle\int_0^tds\, e^{-R_\beta s^{\beta}}\int_0^sds_1\, e^{R_\beta s_1^{\beta}}\int_0^{s_1}ds_2\, e^{-R_\beta s_2^{\beta}}\int_0^{s_2}ds_3\, e^{R_\beta s_3^{\beta}}}{\displaystyle\biggl[\int_0^tds\, e^{-R_\beta s^{\beta}}\int_0^sds_1\, e^{R_\beta s_1^{\beta}}\biggr]^2}.
\end{equation}
\end{widetext}

In contrast to the exact result in the short-time regime, $\kappa_\beta(t)\approx 4$, the $\hat{q}_\beta^{(1)}$-approximation leads to the value $\frac{8}{3}\approx2.667$, thus failing to give the correct description in the short-time regime, however, in the long-time regime the approximation agrees well with exact time dependence. 

\subsection{The velocity autocorrelation function}

The non stationary fluctuations of sBm induce a relaxation of the propulsion autocorrelation function (PACF) $\langle\hat{\boldsymbol{v}}(t)\cdot\hat{\boldsymbol{v}}(0)\rangle$ in the form of a stretched exponential, i.e.,
\begin{equation}\label{VACF}
\langle\hat{\boldsymbol{v}}(t)\cdot\hat{\boldsymbol{v}}(0)\rangle=e^{-R_\beta t^{\beta}},
\end{equation}
which measures the persistence of active motion induced by sBm. Eq.~ \eqref{VACF} was obtained by noticing that $\langle\hat{\boldsymbol{v}}(t)\cdot\hat{\boldsymbol{v}}(0)\rangle=\bigl\langle\cos\bigl(\varphi(t)-\varphi(0)\bigr)\bigr\rangle$ and use of $\Phi(\varphi,t\vert \varphi^\prime,0)=\frac{1}{2\pi}\sum_{n=0}^\infty e^{-R_\beta t^\beta n^2}e^{in(\varphi-\varphi^{\prime})}$, solution of Eq.~\eqref{Phi} with initial distribution $\Phi(\varphi,0\vert \varphi(0),0)=\delta(\varphi-\varphi(0))$. In the case $\beta=1$, the simple exponential relaxation is recovered and  $R_1$ is recognized with the rotational diffusion of standard active Brownian motion, that defines the persistence time $\tau_1=\int_0^\infty dt\, e^{-R_1 t}= R_1^{-1}$. For $\beta\neq1$ we have, analogously
$\tau_\beta=\int_0^\infty dt\, e^{-R_\beta t^{\beta}}=\Gamma\bigl(1+\beta^{-1}\bigr)R_\beta^{-\beta^{-1}}$, this decreases monotonically with $\beta$ being larger than $R_1^{-1}$ for $0<\beta<1$ and smaller for $1<\beta<2$, implying the origin of the origin of the anomalous transport observed.

\begin{figure}
\includegraphics[width=\columnwidth,trim=0 0 0 0,clip]{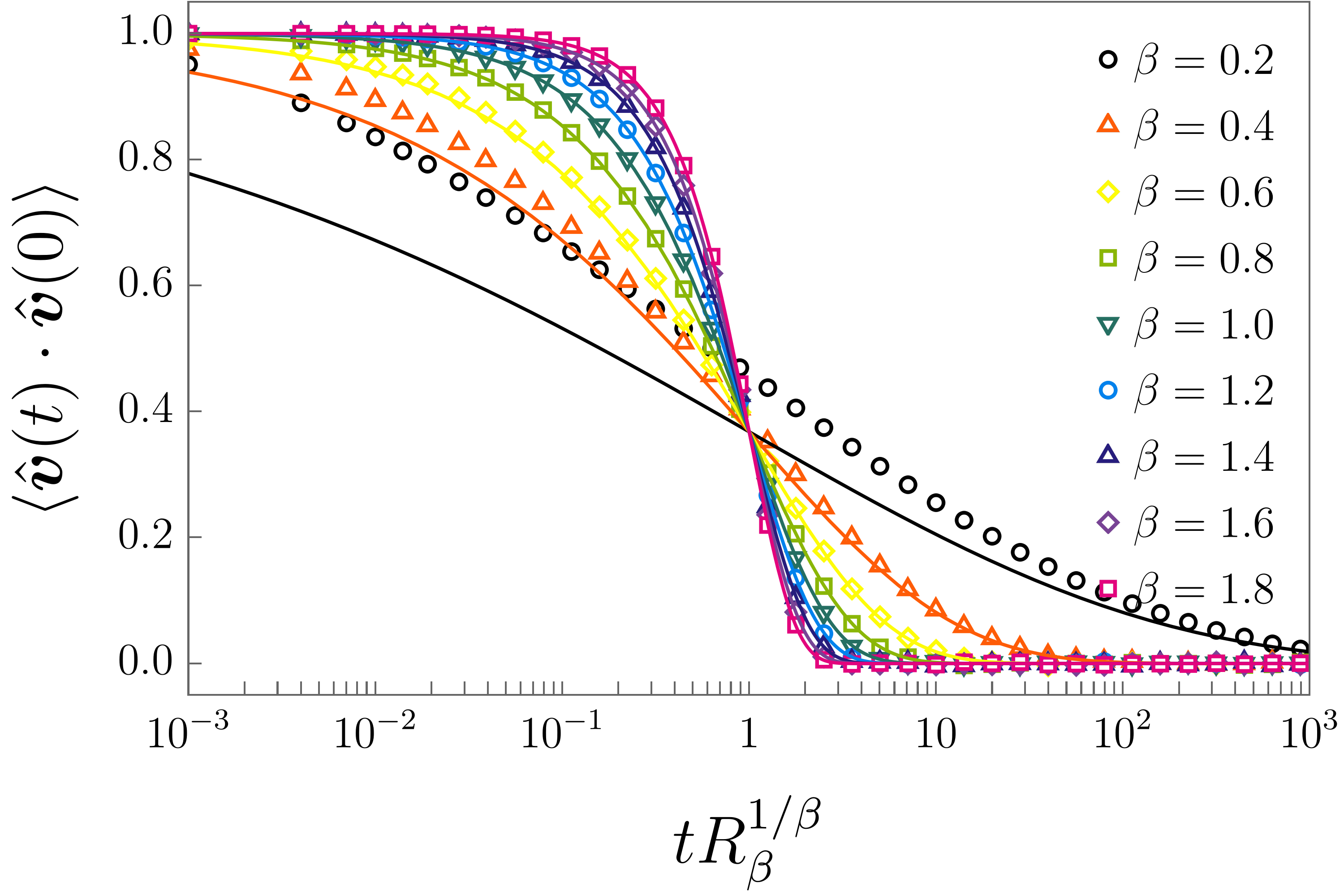}
\caption{Velocity autocorrelation function $\langle\hat{\boldsymbol{v}}(t)\cdot\hat{\boldsymbol{v}}(0)\rangle$. The symbols mark the exact results obtained from numerical simulations for $\beta=0.2$, 0.4, 0.6, 0.8, 1.0, 1.2, 1.4 , 1.6 and 1.8. Solid lines indicate the analytical formula $e^{-R_\beta t^{\beta}}$.}
\label{fig:VACF}
\end{figure}

In Fig.~\ref{fig:VACF} the PACF is shown from the numerical analysis (symbols) as function of the dimensionless time $tR_\beta^{1/\beta}$ for $\beta=0.2$ (dark circles), 0.4 (light up triangles), 0.6 (light diamonds), 0.8 (light squares), 1.0 (dark down triangles), 1.2 (light circles), 1.4 (dark up triangles), 1.6 (dark diamonds) and 1.8 (dark squares); and compared with the analytical stretched exponential.

\section{\label{Conclusions} Final remarks}
Motivated by the anomalous transport properties of tracer particles diffusing in complex media, particularly of tracers that inherits  characteristics of active motion when diffusing in active baths, we propose an extension of active Brownian motion as a model of tracer particles moving in complex media for which the stochastic dynamics of propulsion is driven by scaled Brownian motion, a highly non stationary process paradigmatic of nonequilibrium stochastic dynamics.

A crossover from ballistic transport in the short-time regime, characteristic of active motion, to genuine anomalous diffusion in the long-time regime with exponent $2-\beta$, is observed; $\beta$ is the exponent that characterizes sBm. This has been shown from numerical and analytical calculations. 

We analyzed the intermediate scattering function $\hat{p}_\beta(\boldsymbol{k},t)$, which corresponds to the Fourier transform of the reduced probability density of finding a particle at position $\boldsymbol{x}$ at time $t$ independent of the direction of propulsion $\varphi$, $p_\beta(\boldsymbol{x},t)$. Numerical analysis shows the discrepancy between the results obtained from the analysis of the ensemble of trajectories generated from Eqs.~\eqref{modelo} and the approximation given by Eq.~\eqref{SMoment-Trans}, notwithstanding this, the analytical formula for the MSD \eqref{MSD-Analytical} from such approximation, agrees perfectly from our ensemble analysis. 

The discrepancy is exhibited in the kurtosis of the particle positions distribution $p_\beta(\boldsymbol{x},t)$, mainly in the short-time regime. In this regime, the approximation given by $\hat{q}_\beta^{(1)}$ in Eq.~\eqref{SMoment-Trans}, describe coherent transport describe by the two-dimensional wave equation. This differs from the transport describe by the persistence of active motion.  

\acknowledgments
This work was supported by UNAM-PAPIIT IN112623.

\end{document}